\documentclass[11pt]{article}
\usepackage{jheppub}
\usepackage{amsfonts,ulem,bbold}
\usepackage{amsmath,amssymb, braket}
\newcommand{\gmn}{g_{\mu\nu}}
\newcommand{\bgmn}{\bar g_{\mu\nu}}
\newcommand{\beqn}{\begin{eqnarray}}
\newcommand{\eeqn}{\end{eqnarray}}
\newcommand{\be}{\begin{equation}}
\newcommand{\ee}{\end{equation}}

\newcommand{\sgf}{\sqrt{g^{-1}f}}
\newcommand{\dd}{\mathrm{d}}

\newcommand{\fmn}{f_{\mu\nu}}
\newcommand{\Gmn}{G_{\mu\nu}}
\newcommand{\Mmn}{M_{\mu\nu}}
\newcommand{\bfmn}{\bar f_{\mu\nu}}

\newcommand{\md}{\mathrm{d}}

\title{On Partially Massless Bimetric Gravity}
\author{S.F.~Hassan,}
\author{Angnis~Schmidt-May}
\author{Mikael~von~Strauss}
\affiliation{Department of Physics \& 
        The Oskar Klein Centre,\\
        Stockholm University, AlbaNova University Centre, 
        SE-106 91 Stockholm, Sweden}
\emailAdd{fawad@fysik.su.se}
\emailAdd{angnis.schmidt-may@fysik.su.se}
\emailAdd{mvs@fysik.su.se}

\abstract{We extend the notion of the Higuchi bound and partial
  masslessness to ghost-free nonlinear bimetric theories. This can be
  acheived in a simple way by first considering linear massive spin-2
  perturbations around maximally symmetric background solutions,   
  for which the linear gauge symmetry at the Higuchi bound is
  easily identified. Then, requiring consistency between an appropriate
  subset of these transformations and the dynamical nature of the
  backgrounds, fixes all but one parameter in the bimetric interaction 
  potential. This specifies the theory upto the value of the
  Fierz-Pauli mass and leads to the unique candidate for nonlinear
  partially massless bimetric theory.} 

\keywords{modified gravity, massive gravity, higher spin fields}
\toccontinuoustrue

\begin{document} 
\maketitle
\flushbottom

\section{Introduction}

Fierz and Pauli (FP) obtained a ghost-free linear theory of massive
spin-2 fields in flat space \cite{Pauli:1939xp,Fierz:1939ix}. When
considered in de Sitter backgrounds, the theory has interesting
features associated with the Higuchi bound, $m_\mathrm{FP}^2=
\frac{2}{3}\Lambda$, that determines the mass of the spin-2 state in
terms of the cosmological constant
\cite{Higuchi:1986py,Higuchi:1989gz}. A new gauge symmetry appears at
the bound and eliminates the longitudinal mode of the spin-2 field,
leaving behind only 4 propagating modes \cite{Deser:1983mm,
  Deser:2001pe, Deser:2001us, Deser:2001wx, Deser:2001xr,
  Deser:2004ji, Deser:2006zx}. The resulting theory is the linear
partially massless theory on a de Sitter background.

To better understand the origin and consequences of this symmetry, one
needs to work with a fully dynamical, nonlinear version of the FP
theory. On general grounds, such nonlinear massive spin-2 theories
require working with two metrics, say, $\gmn$ and $\fmn$. But they are
also generically plagued by the Boulware-Deser ghost instability
\cite{Boulware:1972zf,Boulware:1973my}. A major breakthrough in this
field was the work of \cite{deRham:2010ik, deRham:2010kj}. Here, the
authors developed the nonlinear massive gravity for a flat reference
metric $\fmn=\eta_{\mu\nu}$, the dRGT model, and established that it
was ghost free in a certain ``decoupling limit'' analysis 
\cite{ArkaniHamed:2002sp,Creminelli:2005qk} that proved to be very
powerful for the purpose. All subsequent developments in the field are
based on this breakthrough. However, the decoupling limit analysis
cannot show the absence of ghost away from this specific limit. That
the model remained ghost-free at the complete non-linear level was
proven in \cite{Hassan:2011hr}.

As such the dRGT model does not admit de Sitter solutions since it was 
constructed for $\fmn=\eta_{\mu\nu}$, to which the decoupling limit 
analysis is mostly confined. However, the nonlinear analysis of
\cite{Hassan:2011hr} could be used to prove that massive gravity with 
generic non-dynamical reference metric $\fmn$ \cite{Hassan:2011vm} was
ghost-free at the completely non-linear level \cite{Hassan:2011tf}.
Finally the bimetric theory was obtained and proven to be ghost-free
in \cite{Hassan:2011zd,Hassan:2011ea}. This theory admits completely
dynamical de Sitter solutions with massive spin-2 fluctuations around
them. Hence it provides a natural setup for investigating partial
masslessness and the associated symmetries.   

It is now interesting to ask whether there exists a nonlinear
extension of the linear partially massless FP gravity within the
family of ghost-free massive gravity and bimetric theories. In the
context of massive gravity, this question was recently investigated in
\cite{deRham:2012kf}, in a decoupling limit specifically developed for
the de Sitter space. The authors discover the parameter values for
which the St\"uckelberg field that captures the dynamics of the
helicity-zero mode of the graviton is removed to all orders in the
decoupling limit. If this result extends beyond the decoupling limit,
then the theory has a nonlinear gauge symmetry that removes the
helicity-zero excitation, and thus constitutes the nonlinear partially
massless theory of gravity . For a related investigation, see
\cite{Fasiello:2012rw}. 

Here we investigate the problem in the ghost-free bimetric setup where
the backgrounds are not fixed by hand but arise dynamically. A large
class of such theories can satisfy the Higuchi bound and exhibit
linear partial masslessness. It is shown that by demanding consistency
between the dynamical backgrounds and just a subset of the linear
gauge transformations known from the FP theory, one can arrive at a 
unique non-linear candidate for a partially massless bimetric theory. 

Other recent works on partial masslessness, but in somewhat different
setups, include \cite{Iglesias:2010xz,Deser}, as well as
\cite{Hohm:2012vh} in the ``new massive gravity'' framework of
\cite{Bergshoeff:2009aq, Bergshoeff:2009hq}. The issue arises also in 
the context of higher-spin theories (see
e.g.~\cite{Sagnotti:2011qp,Francia:2008hd, Joung:2012rv}).

\section{The Higuchi bound and partial masslessness in linear massive
  gravity} 

The linear Fierz-Pauli equation for a massive graviton on a fixed de
Sitter background reads, 
\be
\label{linmass} 
\bar{\mathcal{E}}_{\mu\nu}^{\rho\sigma}\,h_{\rho\sigma} -\Lambda\left(
h_{\mu\nu}-\frac{1}{2}\bgmn\bar{g}^{\rho\sigma}h_{\rho\sigma}\right) 
+\tfrac{m_{\mathrm{FP}}^2}{2}\,
\left(h_{\mu\nu}-\bar{g}_{\mu\nu}\bar{g}^{\rho\sigma}h_{\rho\sigma}\right)=0\,. 
\ee
$\Lambda$ is the cosmological constant and $m_{FP}$ is the mass
of the spin-2 fluctuation. $\bgmn$ is the de Sitter metric and the
kinetic operator is given by,
\be
\bar{\mathcal{E}}^{\rho\sigma}_{\mu\nu}h_{\rho\sigma} 
=-\tfrac1{2}\Big[\delta^\rho_\mu\delta^\sigma_\nu\bar\nabla^2
+\bar g^{\rho\sigma}\bar\nabla_\mu\bar\nabla_\nu -\delta^\rho_\mu
\bar\nabla^\sigma\bar\nabla_\nu-\delta^\rho_\nu\bar\nabla^\sigma
\bar\nabla_\mu-\bgmn\bar g^{\rho\sigma}\bar\nabla^2 
+\bar g_{\mu\nu}\bar\nabla^\rho\bar\nabla^\sigma
\Big]h_{\rho\sigma}\,.
\label{KO}
\ee
The mass term breaks the symmetry under infinitesimal
reparameterizations. 

It is known that, in this theory, a curious role is played by the
Higuchi bound, 
\be 
m_\mathrm{FP}^2=\tfrac{2}{3}\Lambda \,.  
\label{HB}
\ee 
Above the bound, $m_\mathrm{FP}^2>\frac{2}{3}\Lambda_g$,
(\ref{linmass}) propagates only the five healthy polarizations of the 
massive spin-2 fluctuation. In fact, the mass term is fixed uniquely by
demanding that a sixth ghost mode decouples
\cite{Pauli:1939xp,Fierz:1939ix}. Below the bound, $m_\mathrm{FP}^2<
\frac{2}{3}\Lambda_g$, the helicity zero component of the spin-2 field
becomes a ghost and the theory becomes unstable \cite{Higuchi:1986py,
  Higuchi:1989gz}.  But precisely at the Higuchi bound,
(\ref{linmass}) develops a new gauge symmetry that decouples the 
helicity zero component, leaving only four healthy propagating modes
\cite{Deser:1983mm,Deser:2001pe,Deser:2001us,Deser:2001wx,
  Deser:2001xr,Deser:2004ji}. The theory with this value for the mass
is often referred to as \textit{partially massless}. 

The new linear gauge symmetry that emerges at the Higuchi bound reads 
\cite{Deser:1983mm}, 
\be
\label{gaugesymh}
h_{\mu\nu}\longrightarrow h_{\mu\nu}+ \delta h_{\mu\nu} \qquad
\text{with}\quad \delta h_{\mu\nu}\equiv\left(\nabla_{\mu}
\nabla_{\nu}+\frac{\Lambda}{3}\bar g_{\mu\nu}\right)\xi(x)\,,
\ee 
where, $\xi(x)$ is an arbitrary gauge transformation parameter.
Note that the solutions of $\delta h_{\mu\nu}=0$ give conformal Killing
transformations $\delta x^\mu=\nabla^\mu\xi$ on de Sitter space. These
are excluded since the FP theory has no coordinate invariance.

Understanding the origin of (\ref{gaugesymh}) and partial masslessness
requires a nonlinear version of the FP theory that, furthermore,
treats the background dynamically. This suggests working with the
ghost-free bimetric theory that will be reviewed in the next section
and which will be shown to provide a natural setup for addressing the
issue of partial masslessness.

\section{De Sitter solutions and their perturbations in bimetric theory}

The most general bimetric action with the correct combination of the
kinetic and potential terms that avoids the Boulware-Deser ghost at
the nonlinear level is given by, \cite{Hassan:2011zd, Hassan:2011ea} 
\be\label{action} 
S_{gf}=\int\dd^4 x\left[m_g^2~\sqrt{g}~R(g)+m_f^2~\sqrt{f}~R(f)-
2m^4~\sqrt{g}~\sum_{n=0}^4\beta_n e_n\left(\sgf\right)\right]\,.
\ee
The $e_n(S)$ are the elementary symmetric polynomials of the
eigenvalues of the matrix $S$. The action has seven independent
parameters: the Planck masses $m_g$ and $m_f$, and the five
dimensionless $\beta_n$. The mass scale $m$ is degenerate with these.  
The potential is an extension of the 3-parameter massive gravity
potential \cite{deRham:2010kj} in the formulation of
\cite{Hassan:2011vm}. 
 
The equations of motion for $\gmn$ and $\fmn$, obtained from
(\ref{action}) are given, for example, in \cite{recent} on which the
present section is based. To relate to the Higuchi bound, we are
interested in de Sitter solutions. Generic de Sitter solutions for
both of the bimetric equations are of the type,\footnote{Simply
  requiring both $\gmn$ and $\fmn$ to be dS spacetimes restricts the 
  solutions to such proportional backgrounds.}
\be
\bar{f}_{\mu\nu}=c^2 \bar g_{\mu\nu}\,,
\label{fcg}
\ee
where $c$ is a constant, generically determined in terms of the
parameters of the theory. Indeed, for this ansatz, the $g$ and $f$ 
equations of motion reduce to two copies Einstein's equation, 
\be\label{bgeq}
R_{\mu\nu}(\bar g)-\frac{1}{2}\bar g_{\mu\nu} R(\bar g)+ \Lambda_g\bar
g_{\mu\nu}=0\,,  
\ee 
and a similar equation, again for $\bgmn$, but with a cosmological
constant $\Lambda_f$, where,
\be
\Lambda_g=\frac{m^4}{m_g^2}\left(\beta_0+3c\beta_1+3c^2\beta_2+c^3\beta_3\right)
\,,\quad
\Lambda_f=\frac{m^4}{c^2m_f^2}\left(c\beta_1+3c^2\beta_2+3c^3\beta_3
+c^4\beta_4\right)\,.
\label{Lambdas}
\ee
The consistency of the two equations then requires,
\be
\Lambda_g=\Lambda_f\,.
\label{cosmconstr}
\ee
This provides, in general, a quartic equation that determines $c$ in
terms of the 6 combinations of the seven parameters of the theory,
\be
c=c(\alpha, \beta_n)\,,\qquad {\mathrm{with}},
\qquad\alpha\equiv\frac{m_f}{m_g}\,.
\label{c-fixed}
\ee
For the purpose of these solutions, the relevant regions of the
parameter space are those that lead to a positive Fierz-Pauli mass for
the fluctuation, as given below. 

Let us now consider canonically normalized linear perturbations around
this background, 
\be
g_{\mu\nu}=\bar{g}_{\mu\nu}+\tfrac{1}{m_g} \,\delta g_{\mu\nu}
\,,\qquad
f_{\mu\nu}=c^2\bar{g}_{\mu\nu}+\tfrac{1}{m_f} \,\delta f_{\mu\nu}\,.
\ee
The equations of motion for the perturbations are diagonalized in
terms of a massless mode $\delta\Gmn$ and a massive mode $\delta\Mmn$,
\be
\delta G_{\mu\nu}= \frac{\delta g_{\mu\nu}}{m_g}+\alpha^2\,
\frac{\delta f_{\mu\nu}}{m_f}\,, 
\qquad
\delta M_{\mu\nu}= \frac{1+\alpha^2c^2}{2c}
\left(\frac{\delta f_{\mu\nu}}{m_f}
- c^2 \,\frac{\delta g_{\mu\nu}}{m_g}\right) \,.
\label{dMc}
\ee
The normalizations are explained below. They satisfy the
corresponding equations, 
\begin{align}
&\bar{\mathcal{E}}_{\mu\nu}^{\rho\sigma}\,\delta G_{\rho\sigma}
-\Lambda_g\Big( \delta G_{\mu\nu}-\frac{1}{2}\bar
g_{\mu\nu}\bar{g}^{\rho\sigma}\delta G_{\rho\sigma}\Big) =0\,,
\label{Glineq}  \\
& \bar{\mathcal{E}}_{\mu\nu}^{\rho\sigma}\,\delta M_{\rho\sigma}
-\Lambda_g\Big( \delta M_{\mu\nu}-\frac{1}{2}\bar g_{\mu\nu}
\bar{g}^{\rho\sigma}\delta M_{\rho\sigma}\Big)
+\tfrac{1}{2}m_{\mathrm{FP}}^2\,
\big({\delta M}_{\mu\nu}-\bar{g}_{\mu\nu}\bar{g}^{\rho\sigma}
\delta M_{\rho\sigma}\big) =0\,,
\label{Mlineq}
\end{align}
where $\bar{\mathcal{E}}_{\mu\nu}^{\rho\sigma}$ is given by (\ref{KO})
and the Fierz-Pauli mass of the massive spin-2 mode reads,
\be
m_\mathrm{FP}^2=\frac{m^4}{m_g^2}\left(1+\alpha^{-2}c^{-2}\right)
\left(c\beta_1+2c^2\beta_2+c^3\beta_3\right)\,.
\ee
The normalizations in (\ref{dMc}) are chosen for convenience so that
the mass eigenstates can be regarded as fluctuation of nonlinear modes
\cite{recent}, 
\be
G_{\mu\nu}=g_{\mu\nu}+\alpha^2f_{\mu\nu}\,,\qquad
M_{\mu\nu}=G_{\mu\rho}{\big(\sgf\big)^\rho}_\nu-c\,G_{\mu\nu}\,.
\ee
Equation (\ref{Mlineq}) for the massive spin-2 fluctuation
coincides with the Fierz-Pauli equation (\ref{linmass}). In
particular, at the Higuchi bound, $m_\mathrm{FP}^2=\frac{2}{3}
\Lambda_g$, it has the same extra gauge invariance (\ref{gaugesymh}).
Now, taking the presence of the massless mode $\delta\Gmn$ also into
account, the corresponding symmetry transformations in the bimetric
theory become,
\be
\label{gaugesym}
\delta M_{\mu\nu}\rightarrow \delta M_{\mu\nu}+ 
\Big(\nabla_{\mu}\nabla_{\nu}+\frac{\Lambda}{3}\,
\bar g_{\mu\nu}\Big)\xi(x) \,, \qquad
\delta G_{\mu\nu}\rightarrow\delta G_{\mu\nu}\,.
\ee
While, superficially, this is very similar to Fierz-Pauli massive
gravity, a major difference is that in the bimetric case the
background is dynamical and is not fixed by hand. Demanding
compatibility between a subset of (\ref{gaugesym}) and the dynamical 
nature of the background is powerful enough to uniquely determine 
the partially massless nonlinear bimetric theory. This is explained
below.


\section{Determination of the partially massless bimetric theory}
\label{motivation}

From (\ref{gaugesym}) and (\ref{dMc}), one can easily read off the 
transformations of $\delta\gmn$ and $\delta\fmn$,
\be
\delta\gmn\rightarrow \delta\gmn+ 
a\,\Big(\nabla_{\mu}\nabla_{\nu}+\frac{\Lambda}{3}\,
\bar g_{\mu\nu}\Big)\xi(x) \,, \quad
\delta\fmn\rightarrow\delta\fmn +
b\,\Big(\nabla_{\mu}\nabla_{\nu}+\frac{\Lambda}{3}\,
\bar g_{\mu\nu}\Big)\xi(x)\,,
\label{gaugesym-gf}
\ee
where the constants $a$ and $b$ are given in terms of $\alpha$ 
and $c$. The gauge transformations, being symmetries, are trivial
solutions of the linearized equations of motion.

The crucial point to note is that, for a dynamical field, say $\gmn$,
the split into a background part $\bgmn$ and a fluctuation
$\delta\gmn$ is not unique since, in principle, infinitesimal symmetry 
transformations can be shifted between the two. Hence, one can always
transfer a part of $\delta\gmn$ to the background $\bgmn$ to get, 
\be 
g_{\mu\nu}=\bar{g}_{\mu\nu}+\delta
g_{\mu\nu}=\bar{g}'_{\mu\nu} +\delta g'_{\mu\nu}\,, 
\ee 
where the backgrounds $\bgmn$ and $\bgmn'$ differ by an infinitesimal
symmetry transformation. 
 
In bimetric theory, the two metrics are dynamical and come with their
own equations of motion. If we assume that the theory has a nonlinear
symmetry that manifests itself as (\ref{gaugesym}) around the
backgrounds considered, then this argument tells us that it should be
possible to transfer the infinitesimal transformations
(\ref{gaugesym-gf}) from the fluctuations to the backgrounds $\bgmn$
and $\bfmn$ and end up with new consistent background solutions. But
for a generic gauge parameter $\xi(x)$ in (\ref{gaugesym-gf}), the new
$\bar g'$ and $\bar f'$ are not proportional and hence are not dS
metrics. In this case not much is known about partial masslessness. To
have the calculation under control, we keep to dS backgrounds and
therefore restrict the transformations in (\ref{gaugesym-gf}) to
constant $\xi(x)=\xi_0$.  This subset of (\ref{gaugesym-gf}) is
consistent with the proportional background ansatz and can be used for
the purpose of identifying the PM theory. In this case,
\be 
\delta\gmn\rightarrow \delta\gmn+ a\,\frac{\Lambda}{3}\,\xi\,\bar
g_{\mu\nu}\,, \qquad \delta\fmn\rightarrow\delta\fmn +
b\,\frac{\Lambda}{3}\,\xi\,\bar g_{\mu\nu}\,.
\label{shift-gf}
\ee 
Transferring these from the fluctuations to the backgrounds gives,
\be
\label{linbgtr} \bgmn'=\bgmn+a \frac{\Lambda}{3}\xi\,\bgmn \,,\quad
\bfmn'=\bfmn+b \frac{\Lambda}{3}\xi\,\bgmn 
\ee 
Now, it is obvious that $f'=c'^2\,g'$ with a constant $c'(\xi)$, and
$c'\neq c$. As such $c'$ will differ from $c$ infinitesimally.
However, if we insure that $\bgmn'$ and $\bfmn'$ are exact dS
solutions (not just to linear order in $\delta c=c'-c$), then
perturbations around the new backgrounds are again invariant under
(\ref{gaugesym-gf}) and the process can be repeated to generate a
finite transformation. In other words, we require that transformations
(\ref{linbgtr}) are integrable in the nonlinear theory and can be
iterated to generate finite $\delta c$ (the integrated form of
(\ref{linbgtr}) is given in the next section). 

On the other hand, from the previous section we know that,
generically, the bimetric equations fix $c$ in terms of the parameters
of the theory and hence $c'(\xi)$ cannot be a solution to
(\ref{cosmconstr}). The obvious implication is that, for generic
bimetric parameters, $\bgmn'$ and $\bfmn'$ are not valid background
solutions. This can happen only if the transformations that generate
them are not symmetries of the nonlinear theory. Thus the only
parameter values for which the bimetric theory is consistent with the
transformations (\ref{shift-gf}), are those for which the equation for
$c$ (\ref{cosmconstr}) does not determine $c$ at all! This is the
necessary condition for the existence of a nonlinear partially
massless bimetric theory.

Note that by fixing $\xi(x)$ to be constant, we restrict ourselves to
only part of the gauge group. If in fact the full gauge symmetry
is realized at the nonlinear level, the theory obtained from using
only the subgroup will definitely contain the partially massless
theory invariant under the full symmetry.

Having established a necessary condition for the existence
of the nonlinear partially massless theory, one can easily find the
parameter values that leave the $c$ in the proportional background
ansatz undetermined. The equation (\ref{cosmconstr}) that determines
$c$ can be written as, 
\be
\label{bgconst}
\beta_1+\left(3\beta_2-\alpha^2\beta_0\right)c+\left(3\beta_3-
3\alpha^2\beta_1\right)c^2+\left(\beta_4-3\alpha^2\beta_2\right)c^3
-\alpha^2\beta_3c^4 =0\,.
\ee
Thus the parameter combination for which $c$ remains undetermined is,
\be\label{parch}
\alpha^2\beta_0=3\beta_2\,,\qquad
3\alpha^2\beta_2=\beta_4\,,\qquad
\beta_1=\beta_3=0\,.
\ee
Remarkably, this fixes all but one of the $\beta_n$ which immediately
implies that if there is a nonlinear partially massless theory it has
to be this one. Note that with the choice (\ref{parch}) we have 
\be\label{cosmcon}
m_{\mathrm{FP}}^2=2\frac{m^4}{m_g^2}\left(\alpha^{-2}+c^2\right)
\beta_2=\frac{2}{3}\Lambda_g\,,
\ee
so in particular the massive fluctuation has a mass at the Higuchi bound.

As a side remark we note that the PM action is symmetric under
$\alpha^{-1}\gmn \leftrightarrow \alpha f_{\mu\nu}$ which is a
consequence of $\sqrt{g}~e_n(S)=\sqrt{f}~e_{4-n}(S^{-1})$
\cite{Hassan:2011zd}. While this interchange is a symmetry of
(\ref{action}) whenever the parameters satisfy $\alpha^{-n}\beta_n =
\alpha^{n-4}\beta_{4-n}$, the PM case corresponds to a further
restriction to a single parameter.

The reasoning presented here is simple and straightforward enough that
it can be easily generalized to any number of dimensions and to
theories with multiple spin-2 fields
\cite{Hinterbichler:2012cn,Hassan:2012wt}. 

\section{Nonlinear scaling symmetry}

Taking into account the value for $\Lambda_g$ given in
(\ref{cosmcon}), it is easy to see that for the parameter choice
(\ref{parch}) the background equations (\ref{bgeq}) are invariant
under the simultaneous transformations, 
\be\label{backgsym}
c\longrightarrow c + a\,,\qquad \bar g_{\mu\nu} \longrightarrow
\frac{\alpha^{-2}+c^2}{\alpha^{-2}+(c+a)^2}\bar
g_{\mu\nu}\,,~~~~~~~~a\in\mathbb{R}\,. 
\ee 
This is the nonlinear version of (\ref{linbgtr}) and verifies
that the transformations considered are indeed integrable as
required. The linearized versions of (\ref{backgsym}) read 
\beqn\label{symtrafos} c\longrightarrow c + \delta
c\,,\qquad \bar g_{\mu\nu} \longrightarrow \bar g_{\mu\nu}+\delta \bar
g_{\mu\nu}\equiv\bar g_{\mu\nu}- \frac{2c}{\alpha^{-2}+c^2}\delta
c~\bar g_{\mu\nu}\,. \eeqn Moreover, using $\bar f_{\mu\nu}=c^2 \bar
g_{\mu\nu}$, we find that $\bar f_{\mu\nu}$ transforms as \beqn \bar
f_{\mu\nu}\longrightarrow \bar f_{\mu\nu}+\delta \bar f_{\mu\nu}\equiv
\bar f_{\mu\nu} +2c \delta c~\bar g_{\mu\nu}+c^2\delta \bar
g_{\mu\nu}\,. \eeqn We will now reverse the arguments that were given
in section \ref{motivation} and translate these scalings of the
backgrounds into transformations of the fluctuations in order to see
if we can re-arrive at the $\xi=$const.~version of the transformation
(\ref{gaugesym}) for the massive fluctuation in de Sitter space. For
this we identify $\delta 
f_{\mu\nu}=m_f\delta\bar f_{\mu\nu}$ and $\delta
g_{\mu\nu}=m_g\delta\bar g_{\mu\nu}$. Then we see that the massless
and massive fluctuation transform as \beqn \delta
G_{\mu\nu}\longrightarrow\delta G_{\mu\nu}\,,\qquad \delta
M_{\mu\nu}\longrightarrow \delta
M_{\mu\nu}+\frac{\Lambda_g}{3}\frac{m_f^2}{\beta_2m^4}\bar
g_{\mu\nu}\,\delta c\,, \eeqn where we have used (\ref{cosmcon}) to
express the transformation of $\delta M_{\mu\nu}$ in terms of
$\Lambda_g$. As required, the symmetry transformation
(\ref{symtrafos}) leaves the massless fluctuation
invariant\footnote{Note that, in addition, the background
  $\bar{G}_{\mu\nu}=(1+\alpha^2c^2)\bar{g}_{\mu\nu}$ of the nonlinear
  massless field $G_{\mu\nu}=g_{\mu\nu}+\alpha^2f_{\mu\nu}$ is
  invariant under the full nonlinear scaling symmetry
  (\ref{backgsym}).} while the massive fluctuation transforms as under
a scaling. The transformation of $ \delta M_{\mu\nu}$ is precisely of
the form (\ref{gaugesym}) with
$\xi=\frac{m_f^2}{\beta_2m^4}\delta c$. Thus we have reproduced the
linearized symmetry with constant gauge parameter known to be present
at the Higuchi bound through the constant but nonlinear transformation
(\ref{backgsym}).
 
We emphasize once more that all of these conclusions become invalid
once we choose parameters different from (\ref{parch}) since then the
background equations do determine $c$ and there is no invariance at
the background level. Thus the unique candidate for a nonlinear
partially massless theory is the one specified by (\ref{parch}).

\section{Summary and Discussions}\label{disc}

Ghost-free bimetric theories admit de Sitter backgrounds with
Fierz-Pauli massive spin-2 excitations. A large subclass of these
theories can easily satisfy the Higuchi bound and exhibit partial
masslessness at the linear level. However, using simple symmetry
arguments, we have shown that it is possible to identify a unique
theory in this multi-parameter space as the only possible candidate
for a non-linear PM theory. The construction involves showing that (i)
the known linear PM transformations contain a simple de Sitter
preserving subgroup generated by constant gauge parameter $\xi_0$,
and that (ii) in any non-linear PM setup it must be possible to
integrate such linear dS-preserving transformations to a non-linear
one. This is the necessary condition that must be satisfied by any
non-linear PM theory, if such a theory exists. In the bimetric setup
in 4-dimensions, this singles out a unique theory specified by
\eqref{parch} as the only possible candidate for a non-linear PM
theory. While the construction narrows down the search to a unique
candidate, it does not prove that it actually has a PM symmetry. This
may or may not be the case and the proof is beyond the scope of this
paper. 

While this paper focuses on bimetric theories, our logic and
construction is general and can be applied to any nonlinear setup that
accommodates linear PM symmetry around dS backgrounds to identify the
potential non-linear PM candidates.

In \cite{deRham:2012kf}, the authors identify a specific nonlinear
massive gravity which, in a decoupling limit adapted to dS spacetimes,
shows a non-linear PM symmetry. To compare results, we need to take a
massive gravity limit of \eqref{parch} by setting $m_f=\infty$ (or
$\alpha=\infty$) keeping $\beta_2$ fixed. This gives $\beta_0=0$ and
$\beta_4=\alpha^2\beta_4'$ for a finite $\beta_4'$, agreeing with the
parameters for which \cite{deRham:2012kf} finds an extra PM symmetry
in the decoupling limit \footnote{Comparison with the theory found in
  \cite{deRham:2012kf} requires some care. The massive gravity action
  presented there contains the original dRGT potential which is
  constructed with $\alpha_1=0$ so that the $\gmn=\fmn$ solutions have
  zero cosmological constant. Using this action one reads off
  $\beta_1=\beta_3=0$, $\beta_2=-\frac{1}{2}$ and
  $\beta_0=\frac{3}{2}\neq 0$ (for example, using the relations in
  appendix B of \cite{vonStrauss:2011mq}). However, to find the PM
  symmetry, \cite{deRham:2012kf} considers perturbations around
  $\gmn=\fmn$ backgrounds with a de Sitter $\fmn$. The correct action
  corresponding to this must have $\alpha_1\neq 0$ such that
  $\beta_0=0$, consistent with the massive gravity limit of the theory
  found here.}, although, in general the two formalisms lead to
different predictions, as described below.

An important question that remains unanswered here is, if the
candidate non-linear theory can really have the full PM gauge
symmetry, beyond the dS preserving subset used here.  Recent evidence
for and against this will be discussed below. Here we point out that
the theory specified by \eqref{parch} definitely has an extra
nonlinear gauge symmetry for one class of non-proportional
backgrounds. Consider non-proportional homogeneous and isotropic
backgrounds parameterized by three functions $a(t)$, $Y(t)$ and $X(t)$
\cite{vonStrauss:2011mq},
\be 
\gmn \md x^\mu \md x^\nu=-\md t^2 + a^2(t) \md\vec x^2 \,,\qquad \fmn
\md x^\mu \md x^\nu=-X^2(t)\md t^2 + Y^2(t) \md \vec x^2 
\label{cs}
\ee 
The bimetric equations can in general be solved to determine the three
functions. The fluctuations around these backgrounds do not have the
Fierz-Pauli form. However, for the parameter values specified by
\eqref{parch}, and in the absence of sources, the equation that
determines $\Upsilon=Y/a$ disappears, as can be verified from
\cite{vonStrauss:2011mq}. Hence the theory leaves one of the three
functions in \eqref{cs} undetermined. This implies a nonlinear gauge
symmetry of the cosmological metrics \eqref{cs} in the candidate PM
theory, well beyond the linear PM symmetry \eqref{gaugesym-gf} around
proportional backgrounds, albeit only for a cosmological gauge
parameter.  It is also interesting to understand the partially
massless bimetric theory in connection with the self-protection
mechanism discussed in
\cite{Grisa:2009yy,Berkhahn:2010hc,Berkhahn:2011jh,Berkhahn:2011hb}.

Since this work first appeared, several papers have addressed the
issue of PM symmetry in nonlinear massive gravity and in bimetric
theory, providing arguments both in favour and against it. Below we
briefly comment on these developments. \cite{Hassan:2012rq} provides
evidence for the strength of the method used here by applying it to
general dimensions. It finds that PM theories cannot exist for $d>4$
in a 2-derivative theory, consistent with perturbative reasoning. This
differs from the prediction in \cite{deRham:2012kf} which is not
dimension sensitive. Furthermore, \cite{Hassan:2013pca} finds evidence
for an extra gauge symmetry in the candidate PM theory by partially
solving the bimetric equations perturbatively in powers of curvatures
of $\gmn$. Then, at the 4-derivative level, the PM bimetric equations
coincide with the conformal gravity equation of motion which
propagates 6, rather than 7, modes. Then, to this order, the PM
symmetry is related to the Weyl scaling of $\gmn$. The relation to
conformal gravity already shows the special status of the theory
identified by \eqref{parch}, irrespective of the final fate of the PM
symmetry.

On the other hand, \cite{deRham:2013wv, Deser:2013uy} argue against
the existence of PM symmetry in massive gravity. Specifically,
\cite{deRham:2013wv} cannot find a PM symmetry in the massive gravity
action (which it takes as \eqref{action} without the $\sqrt{f}R(f)$
term) while \cite{Deser:2013uy} cannot find the associated Binachi
identities. While these studies are limited to massive gravity, one
may wonder if the claimed absence of a PM massive gravity could
already rule out a PM bimetric theory. If the PM bimetric theory
happens to have a well behaved massive gravity limit, then one expects
its PM symmetry to survive in the limit. If so, the absence of a PM
massive gravity would also imply the absence of a PM bimetric
theory. While this is not ruled out at this stage, there are
indications that the massive gravity limit of the PM bimetric theory
is not completely well behaved in the sense that not all bimetric
solutions have a massive gravity limit.

Massive gravity is obtained by setting $m_f=\infty$ in the bimetric
equations, then the $\gmn$ equation is unchanged, while the $\fmn$
equation reduces to an Einstein equation,
\be
m_g^2G_{\mu\nu}(g)+V_{\mu\nu}^g=0\,,\qquad G_{\mu\nu}(f)+\beta_4'
f_{\mu\nu}=0\,, 
\ee 
where $G_{\mu\nu}$ denotes the Einstein tensor and $\beta_4'=
\beta_4/m_f^2$ is fixed. These two equations can no longer be obtained
from an action principle. If the PM gauge transformation of bimetric
theory depends non-trivially on $m_f$, it will not survive on setting
$m_f=\infty$. The bimetric solutions generated by such transformations
are the ones that do not have a massive gravity limit. 
Using the perturbative solution for $\fmn$ in powers of
curvatures of $\gmn$, found in \cite{Hassan:2013pca}, one can check
that the transformation of $\fmn$ induced by a Weyl scaling of 
$\gmn$ contains terms that diverge for $m_f=\infty$, but which vanish
for a dS (or Einstein) $\gmn$. This leads one to expect that 
the gauge symmetry of the candidate PM bimetric theory is lost in the 
massive gravity, except around dS backgrounds.

Another difference can be seen at the level of classical solutions. In
massive gravity, where $\fmn$ is always an Einstein metric, the
fluctuation spectrum around $\gmn\propto\fmn$ can have a PM
symmetry. But, it is also possible to find a $\gmn$ background that  
is not an Einstein metric and around these PM symmetry is evidently
lost. On the contrary, in the candidate bimetric PM theory, 
one can show that whenever $\fmn$ is an Einstein metric then $\gmn$ is
{\it necessarily a proportional Einstein metric} in which case the
linear spectrum always has a PM symmetry. Thus, for any {\it finite} $m_f$,
it is impossible to find PM violating massive gravity type backgrounds
where only $\fmn$ (but not $\gmn$), is an Einstein metric. This
constraint disappears precisely for $m_f=\infty$, in which case PM
violating backgrounds become allowed classical solutions. Finally, it
is likely that a PM symmetry exists only on-shell and not at the level
of the action. The above arguments indicate that the massive gravity
investigations in \cite{deRham:2013wv, Deser:2013uy} do not
necessarily rule out a PM bimetric theory.

\vspace{.3cm}

\noindent
{\bf Acknowledgments:} We would like to thank Augusto Sagnotti for
asking the questions that led us to the present investigation.

\end{document}